\documentclass[]{aastex631}

\newcommand{\lbparticles}{\texttt{lbparticles}}

\usepackage{amsmath}

\shorttitle{Lynden-Bell Orbits}
\shortauthors{Forbes et al.}

\graphicspath{{./}{figures/}}

\begin{document}

\title{Semi-analytic Orbits: A Practical Implementation of Lynden-Bell's Planar Orbits and Extension to Vertical Oscillations}

\author[0000-0002-1975-4449]{John C. Forbes}
\affil{School of Physical and Chemical Sciences--Te Kura Mat\=u, University of Canterbury, Christchurch 8140, New Zealand}

\author[0000-0003-3257-4490]{Michele T. Bannister}
\affil{School of Physical and Chemical Sciences--Te Kura Mat\=u, University of Canterbury, Christchurch 8140, New Zealand}

\author[0009-0008-0355-5809]{Angus Forrest}
\affil{School of Physical and Chemical Sciences--Te Kura Mat\=u, University of Canterbury, Christchurch 8140, New Zealand}

\author[0009-0004-3142-3898]{Ian DSouza}
\affil{School of Physical and Chemical Sciences--Te Kura Mat\=u, University of Canterbury, Christchurch 8140, New Zealand}

\author[0009-0001-1692-4676]{Jack Patterson}
\affil{School of Physical and Chemical Sciences--Te Kura Mat\=u, University of Canterbury, Christchurch 8140, New Zealand}

\correspondingauthor{John C. Forbes}
\email{john.forbes@canterbury.ac.nz}

\begin{abstract}
We present a practical implementation of the perturbation theory derived by Lynden-Bell (2015) for describing, to arbitrary precision, the orbit of a particle in an arbitrary spherically-symmetric potential. Our implementation corrects minor but important errors in the initial derivation, and extends the formalism in two ways. First, a numerical method is developed to efficiently and precisely solve the analogue to the Kepler problem, and second, a method is introduced to track the particle's vertical oscillations about an axisymmetric disk, even when the vertical oscillation frequency varies with radius. While not as flexible as numerical integration, this method guarantees conservation of energy, angular momentum, and related quantities, and may be used to evaluate a particle's position and velocity in constant time. Our implementation is written in Python and is pip installable as the package \lbparticles{}.
\end{abstract}

\section{Introduction} \label{sec:intro}

The position and velocity of a classical particle orbiting in a potential defined by a central force is of fundamental importance in astrophysics, with its study dating back to the first attempts to mathematically describe the motion of the Solar System's planets. 
In planetary systems, both in the Solar System and around other stars, the potential may be closely approximated by a $1/r$ potential: i.e. a $1/r^2$ force directed from the orbiting object to the star. 
Other objects, e.g. giant planets, and effects, e.g. those of General Relativity, complicate the problem, and have been addressed by numerically integrating the equations of motion of the planets (see \texttt{ASSIST}\footnote{\url{https://assist.readthedocs.io/en/latest/}} for example).
This is particularly important for applications where high levels of precision are necessary, such as interpreting pulsar arrival times, astrometric observations of small bodies in the Solar System, the long-term stability of the Solar System, and transit timing variations in exoplanetary systems. 

In a Galaxy, the potential is closer to logarithmic (yielding a flat rotation curve). 
The resulting orbits are more difficult to approximate analytically, since the orbits do not close and cannot be described by ellipses. 
Two factors make numerical integration quite straightforward with off-the-shelf techniques \citep[e.g.][]{bovy_Galpy_2015}: the computational power of modern computers, and the relatively small number of dynamical times that the Galaxy has experienced near the Solar orbit ($\sim 13.7~\mathrm{Gyr} / 240~\mathrm{Myr} \approx 57$).
Modern studies of the motion of objects in the Galaxy's potential almost universally just integrate the particles directly\citep[e.g.][]{zhang_Search_2022, koposov_Probing_2023, el-badry_Transiting_2023, miret-roig_Insights_2024}. 

\begin{figure}
    \centering
    \includegraphics[width=0.75\linewidth]{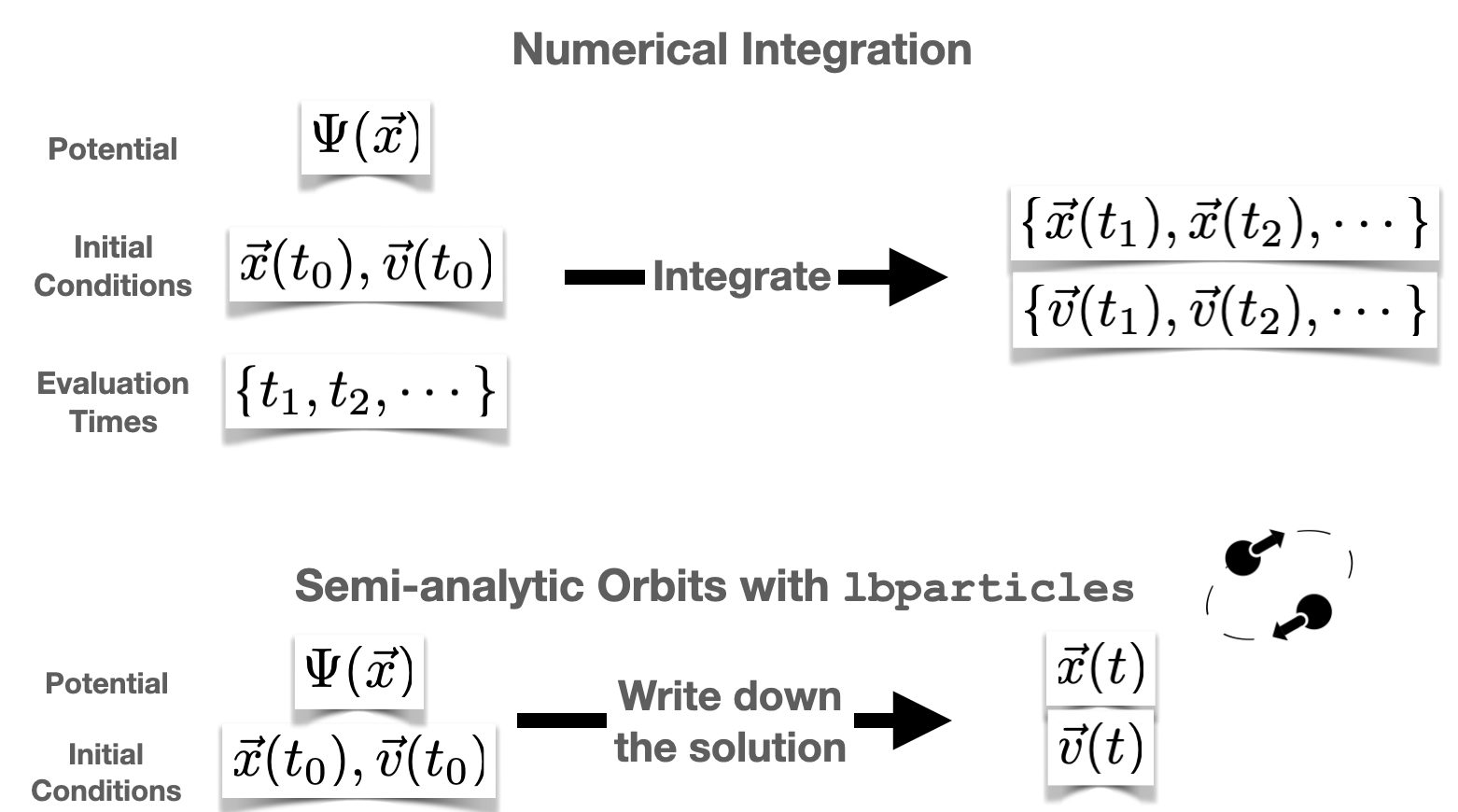}
    \caption{Schematic illustrating the difference between numerical integrations of orbits, and semi-analytic descriptions of orbits. Numerical integration requires a list of evaluation times, and storing the locations and velocities of the particle at each time. In contrast \lbparticles{}\ and related approaches yield functions of $t$ that can be evaluated at any time. We use the term semi-analytic to emphasize that the functions do have a numerical component, namely the mapping between $t$ and the angles $\chi$ and $\eta$. Once these angles are known, the values of $\vec{x}$ and $\vec{v}$ can be computed analytically (see section \ref{sec:lb_formalism}).}
    \label{fig:schematic}
\end{figure}

Analytic approximations to orbits in Galaxy-like potentials have an advantage over numerical integrations when the number of orbits is large, or the necessary time resolution of an object's path is high or unknown prior to the integration, both of which may make computing and storing results from the integrations cumbersome (see Figure \ref{fig:schematic}).
The most prominent of these analytic approximations is the epicyclic approximation --- as described in \citet{binney_Galactic_2008}, developed by \citet{lindblad_Dispersion_1958}. 
These approximations have been useful in deriving the rate at which giant molecular clouds dynamically heat stars in the disk \citep{lacey_influence_1984}, features in the density of stellar streams \citep{kupper_Structure_2008}, and features in galactic tidal streams \citep{struck_Applying_2010}. 
The description of the orbit in terms of radial and vertical energy can also be helpful in the construction and use of analytic distribution functions \citep{dehnen_Simple_1999, daniel_Constraints_2018}. 

The Milky Way has a particularly dynamically cold disk, both relative to our nearest neighbors \citep{dorman_Clear_2015,dalcanton_Panchromatic_2023}, and across cosmic time \citep{kassin_epoch_2012}. 
Despite this, the velocity dispersion of stars in the Solar neighborhood ranges up to 50 km s$^{-1}$ for older stars \citep{holmberg_genevacopenhagen_2009}.
This means that even in the dynamically-cold Solar neighborhood, many stars are far enough from ``nearly-circular'' orbits that the epicyclic approximation becomes inaccurate for the trajectory of any particular star. 
Several authors have therefore attempted to improve on the accuracy of the underlying approximation to extend its validity. 
\citet{kalnajs_Better_1979} noted a simple modification along these lines, which was explored more extensively by \citet{dehnen_Approximating_1999}. 
Similar approaches include precessing ellipses presented by \citet{struck_Simple_2006}, spirographs \citep{adams_Orbits_2005a}, and transformations \citep{lynden-bell_Analytic_2008, lynden-bell_Analytic_2010, binney_Actions_2012, lynden-bell_Bound_2015, mackereth_Fast_2018}.

In this work we provide an accurate analytic approach for a particle orbiting in a Galactic potential, with relevance for both the Solar neighbourhood and a range of other applications.
We focus particularly on \citet{lynden-bell_Bound_2015} (hereafter Lynden-Bell), which includes not only a good first-order theory, but also a full perturbation theory allowing one to in principle find the particle's position and velocity with arbitrary precision in any spherically-symmetric potential.  
In the next section (\S~\ref{sec:lb_formalism}) we provide an overview of Lynden-Bell's method, including minor corrections to the original publication and some deviations from Lyden-Bell's method necessary for practical computations. 
We then extend the method to include vertical oscillations, appropriate for orbits where the self-gravity of a thin disk is important in \S~\ref{sec:vertical}. 
We provide an open-source Python implementation of our improved version of Lynden-Bell's prescription, \lbparticles{}, which is described in more detail in an accompanying JOSS paper.
We provide accuracy benchmarks for \lbparticles{}\ in \S~\ref{sec:results}, and a range of applications in \S~\ref{sec:applications}. 
Throughout, we use right-handed coordinate systems with an origin at the center of the potential, where the $x-y$ (Cartesian) or $r-\phi$ (cylindrical) plane corresponds to the plane of the orbit (in the next section) or the plane of the disk (in subsequent sections).

\section{Lynden-Bell (2015)'s approach and appropriate refinements}
\label{sec:lb_formalism}

As with other analytic models, \citet{lynden-bell_Bound_2015} looks to evaluate the angular position of a particle in its orbit $\phi$ and the time $t$ that has evolved over the course of an orbit. These are computed via the following integrals
\begin{equation}
\label{eq:dphi}
    \phi = \int \frac{h dt}{r^2} = \int \frac{h dr}{r^2 \dot{r}}
\end{equation}
and
\begin{equation}
\label{eq:dt}
    t = \int \frac{dr}{\dot{r}}.
\end{equation}
Here $h=r v_\phi$ is the angular momentum of the particle, where $v_\phi = r\dot{\phi}$ is the tangential velocity and throughout a dot refers to a time derivative. In both cases an integral over time has been transformed to an integral over $r$. These integrals will be further transformed to make them tractable, inspired by Kepler's transformation of $\mathcal{U}=1/r$\footnote{$\mathcal{U}$ is used here to avoid confusion with Lynden-Bell's $u$ to be introduced later}. We will now focus on each of these integrals in turn.

\subsection{Orbit Shape: $\phi$ as a function of phase}
The radial motion of a particle can be understood as a consequence of conservation of energy and angular momentum, $\epsilon = (1/2) \dot{r}^2 + (1/2) h^2/r^2 - \psi(r)$. Here $\psi(r)$ is the gravitational potential\footnote{Note that the sign convention of $\psi(r)$ is negative, so that ``falling down the potential well'' or losing potential energy corresponds to a larger value of $\psi(r)$.}, and $\epsilon$ is the particle's energy. The advantage of transforming the integrals to depend on $r$ is that everything in the integrand is now either a constant or has an explicit dependence on $r$ only. The next step is to choose a transformation of $r$ such that this integrand is nearly one divided by the square root of a quadratic. Lynden-Bell adopts $u=r^{-k}$, where $k$ is a constant for any given orbit. The integrand in Equation \ref{eq:dphi} then becomes $-du / \sqrt{s(u)}$, where
\begin{equation}
    s(u) = (2\epsilon + 2\psi - h^2/r^2) (r^2/h^2) (r du/dr)^2.
\end{equation}
To factor out the quadratic component, Lynden-Bell defines
\begin{equation}
    s_q(u) = m_0^2 (u_p - u)(u-u_a) = s(u)(1+w)^2,
\end{equation}
where $w$ is a perturbation to be computed, $m_0$ is a constant to be determined, and $u_a$ and $u_p$ are the values of $u$ at apo- and peri-center respectively. The $r$ values of peri- and apo-center, $r_p$ and $r_a$, are determined by finding the roots of $\dot{r}(r)=0$, which generally need to be computed numerically. These roots are quick to find with Halley's method \citep{halley_Methodus_1694} provided that the first and second radial derivatives of the potential are known.

By matching $s_q$, and its derivatives, to $s$ itself at peri- and apo-center, Lynden-Bell fixes $m_0$ and $k$ respectively as
\begin{equation}
    m_0^2 = 2 k \frac{1 + \mathcal{R}_p}{1-(r_p/r_a)^k}
\end{equation}
and
\begin{equation}
    k = \frac{\ln\left[-(\mathcal{R}_a+1) / (\mathcal{R}_p+1)\right]}{\ln(r_a/r_p)}
\end{equation}
where $\mathcal{R}_a$ and $\mathcal{R}_p$ are the ratio of gravitational to centrifugal force evaluated at apo- and pericenter respectively. At any given radius $r$,
\begin{equation}
    \mathcal{R}(r) = \frac{d\psi}{dr} \frac{r^3}{h^2}.
\end{equation}
For the shape of the orbit, the final substitution is
\begin{equation}
    r^{-k} = u = \bar{u}(1 + e \cos \eta ),
\end{equation}
where $\bar{u}=(u_a+u_p)/2$, and the eccentricity $e$ is 
\begin{equation}
    e = \frac{u_p-u_a}{u_a+u_p}.
\end{equation}
The particle therefore undergoes a full radial oscillation once $\eta$ has elapsed by $2\pi$. With this substitution,
\begin{equation}
\label{eq:phi}
    \phi = m_0^{-1} \int_0^\eta (1 + e^2 \sin^2(\eta') W) d\eta',
\end{equation}
where the correction function $w$ has been replaced by $W$, defined via $w=(u_p-u)(u-u_a)\bar{u}^{-2}W(u)$. Lynden-Bell adopts the convention that $t=\phi=\eta=0$ at pericenter, which sets the bounds and constants of integration of the integrals.

Equation \ref{eq:phi} essentially says that $\phi = \eta/m_0$ plus a correction proportional to $e^2/m_0$ and related to the deviation of $s(u)$ from quadratic, as encapsulated by $W$. To capture this deviation and keep the integral analytic, Lynden-Bell expands $W$ in a cosine series, 
\begin{equation}
\label{eq:Wseries}
    W(\eta) = (1/2) W_0 + \sum_{n=1}^{N_\mathrm{shape}-1} W_n \cos (n\eta).
\end{equation}
As $N_\mathrm{shape} \rightarrow \infty$, $W$ fully encapsulates arbitrary deviations of $s_q$ from $s$, but practically the series needs to be truncated at a finite value and the $W_n$ need to be estimated. To find the $W_n$, one first evaluates
\begin{equation}
\label{eq:Weval}
    W(\eta) = \frac{\bar{u}^2}{(u_p-u)(u-u_a)} \left( \sqrt{\frac{s_q(u)}{s(u)}} - 1 \right)
\end{equation}
at the zeroes of $\cos(N_\mathrm{shape} \eta)$, i.e. $\eta_{[j]} = (\pi/2 + j \pi)/N_\mathrm{shape}$ for $j=0,...,N_\mathrm{shape}-1$. We then define a matrix
\begin{equation}
\label{eq:Cmatrix}
C = \begin{bmatrix}
0.5 & \cos(\eta_{[0]}) & \cos(2\eta_{[0]}) & \cdots & \cos((N_\mathrm{shape}-1) \eta_{[0]})\\
0.5 & \cos(\eta_{[1]}) & \cos(2\eta_{[1]}) & \cdots & \cos((N_\mathrm{shape}-1) \eta_{[1]})\\
\vdots & \vdots & \vdots & \ddots & \vdots \\
0.5 & \cos(\eta_{[N_\mathrm{shape}-1]}) & \cos(2\eta_{[N_\mathrm{shape}-1]}) & \cdots & \cos((N_\mathrm{shape}-1) \eta_{[N_\mathrm{shape}-1]})
\end{bmatrix}.
\end{equation}
The $W_n$ are then found by solving the matrix equation $W(\eta_{[j]}) = C_{jn} W_{n}$, where the right-hand side uses Einstein summation. We are just setting the right hand sides of Equations \ref{eq:Wseries} and \ref{eq:Weval} equal to each other at the $\eta_{[j]}$ and solving the resultant linear equation for the unknown $W_n$. The inverse of $C$ can be precomputed for all potentials and orbits, since $C$ only depends on $N_\mathrm{shape}$. The $W_n$ can then be found with a matrix multiplication once the $W(\eta_{[j]}$), which will vary from orbit to orbit even in the same potential, have been evaluated.

Once the $W_n$ are evaluated, $\phi$ can immediately be evaluated as 
\begin{equation}
    \phi = m_0^{-1}\left( \eta + (1/4) e^2 \left( (W_0 - W_2)\eta + (W_1 - W_3) \sin \eta + \sum_{k=2}^{N_\mathrm{shape}+1} (-W_{k-2} + 2W_k - W_{k+2})  \frac{\sin(k\eta)}{k} \right) \right)
\end{equation}
via the analytic integration of Equation \ref{eq:phi}. Several features of this equation require explanation. The factors of the form $-W_{k-2} + 2W_k - W_{k+2}$ arise via 2 applications of the formula $\cos(n x) = 2 \cos x \cos((n-1)x) + \cos((n-2)x)$, which is used to express the $\sin^2\eta$ factor in terms of cosines. 
The \citet{lynden-bell_Bound_2015} version of this equation (his 4.34) has an incorrect factor of 2 multiplying $W_1$. Finally, $W_n =0$ for $n\ge N_\mathrm{shape}$ -- terms in this sum that include these $W_n$ {\em are} necessary in the equation above, which is why the sum over $k$ goes to $N_\mathrm{shape}+1$, but some of the $W_n$ they contain will be zero. because Equation \ref{eq:Wseries} is truncated at $n=N_\mathrm{shape}-1$.

\subsection{Particle motion: $t$ as a function of phase}
\label{sec:t}
We next turn to solving Equation \ref{eq:dt}. In the end the integral will not be analytic, but can be precomputed before initializing any particles. This strategy will also be useful for computing the vertical motion of particles in the following section.

Lynden-Bell uses the following transform\footnote{We note that Lynden-Bell's version of this equation prior to transforming to an integral over $\chi$ (his 2.18) contains an incorrect factor of $\l^2$ that should only be introduced after transforming from $u$ to $\chi$, but this error disappears and his equation 2.20 is identical to our Equation \ref{eq:tint}.} to tackle the time integral:
\begin{equation}
\label{eq:U}
    U = 1/u = r^k = \bar{U}(1-e\cos\chi).
\end{equation}
Under this transform, the time integral becomes
\begin{equation}
\label{eq:tint}
    t = \frac{l^2}{h m_0 (1-e^2)^{\nu_k+(1/2)}} \int_0^\chi (1-e\cos\chi')^{\nu_k} (1+w) d\chi',
\end{equation}
where we have introduced $l=\bar{u}^{-1/k}$, and $\nu_k = 2/k - 1$, and where as before $w=\sqrt{s_q(u)/s(u)} - 1$. 
Following a similar strategy as in the previous subsection, Lynden-Bell expands
\begin{equation}
\label{eq:littlew}
    w = e^2 \sin^2\chi\left( (1/2) w_0 + \sum_{n=1}^{N_\mathrm{time}-1} w_n \cos(n\chi) \right).
\end{equation}
As in the previous subsection we can solve a similar linear equation to determine the $w_n$:
\begin{equation}
\begin{bmatrix}
w(\chi_{[0]}) / \sin^2\chi_{[0]}\\
w(\chi_{[1]}) / \sin^2\chi_{[1]}\\
\vdots\\
w(\chi_{[N_\mathrm{time}-1]}) / \sin^2\chi_{[N_\mathrm{time}-1]}
\end{bmatrix}=
    e^2 \begin{bmatrix}
0.5 & \cos(\chi_{[0]}) & \cos(2\chi{[0]}) & \cdots & \cos((N_\mathrm{time}-1) \chi_{[0]})\\
0.5 & \cos(\chi_{[1]}) & \cos(2\chi_{[1]}) & \cdots & \cos((N_\mathrm{time}-1) \chi_{[1]})\\
\vdots & \vdots & \vdots & \ddots & \vdots \\
0.5 & \cos(\chi_{[N_\mathrm{time}-1]}) & \cos(2\chi_{[N_\mathrm{time}-1]}) & \cdots & \cos((N_\mathrm{time}-1) \chi_{[N_\mathrm{time}-1]})
\end{bmatrix}
\begin{bmatrix}
w_0\\
w_1\\
\vdots\\
w_{N_\mathrm{time}-1}
\end{bmatrix},
\end{equation}
where the $\chi_{[j]} = (\pi/2 + j\pi)/N_\mathrm{time}$ for $j=0,\cdots,N_{\mathrm{time}}-1$ are the zeros of $\cos(N_\mathrm{time}\chi)$. The left-hand side contains the true values of $w/\sin^2\chi$ at these zeros as evaluated by computing $s$ and $s_q$. As before $w_n=0$ for $n\ge N_\mathrm{time}$ by construction.

Lynden-Bell explains that his motivation was to create a tractable expansion for $T_r = t(\chi=2\pi)$, the period of a single radial oscillation. In this form, $T_r$ can be computed with a series of Legendre polynomials. However to compute the particle's trajectory during an orbit, Lynden-Bell's proposed method was not usable to arbitrary accuracy, and so may not be suitable for some applications. We still retain the expansion used by Lynden-Bell because this allows us to split the integral in Equation \ref{eq:tint} into the following array of integrals that, once computed, will work for any particle.
\begin{equation}
\label{eq:tauint}
        \tau_{n}(\chi; k, e) = \int_0^\chi (1-e\cos\chi')^{\nu_k} \cos(n\chi) d\chi'.
\end{equation}
We call these integrals $\tau_n$ to emphasize that they are nondimensional because the pre-factor, which depends on other quantities associated with a particular orbit, has been removed, to be multiplied back in later. For these pre-computed integrals to be useful, we need to be able to estimate their values at arbitrary values of $k$ and $e$, which will vary from orbit to orbit. We proceed by expanding the $\tau_n$ in a 2D Taylor series in $\delta e = e-e_0$ and $\delta \nu_k = \nu_k - \nu_{k,0}$, which we show in Appendix \ref{sec:taylor}.
The result is (Equation \ref{eq:qn})
\begin{equation}
\label{eq:tau_n_chi_k_e}
    \tau_n(\chi;k,e) = \sum_{j=0}^\infty \frac{(-\delta e)^j}{j!} \frac{\Gamma(\nu_{k_0}+1)}{\Gamma(\nu_{k_0}-j+1)} \sum_{m=0}^\infty \frac{(\delta \nu_k)^m}{m!}  \sum_{p=0}^m a_p^{(m)}  T_n(\chi,k_0,e_0;j,p) 
\end{equation}
where the gamma function is the analytic continuation of $\Gamma(z)=\int_0^\infty x^{z-1}e^{-x} dx$. The ratio of gamma functions is just $\nu_k(\nu_k-1)(\nu_k-2)\cdots(\nu_k -j)$. The integrals are encapsulated in the final term, the $T_n$.

Within the package \lbparticles{} we implement a class called \texttt{Precomputer} whose job it is to enumerate the integrals contained within Equation \ref{eq:tau_n_chi_k_e}, namely the $T_n(\chi,k,e;j,m)$ defined in Equation \ref{eq:integrands}. The integrals are evaluated at a series of $\chi$ values from 0 to $2\pi$, with $N_\chi$ evenly-spaced points. Separate integrals need to be evaluated for as many values of $n$, $j$, and $m$ as we would like to include. The time series needs $N_\mathrm{time}+2$ values of $n$, where the extra 2 comes from the factor of $\sin^2\chi$ in Equation \ref{eq:littlew}, while $j$ and $m$, correspond to terms in the 2D Taylor series in $\delta e = e-e_0$ and $\delta \nu_k = \nu_k - \nu_{k,0}$. We include values of $j$ up to $N_e=10$ by default, and values of $m$ up to $N_{\nu_k}=5$ by default.

 \begin{figure}
     \centering
     \includegraphics[width=0.7
     \textwidth]{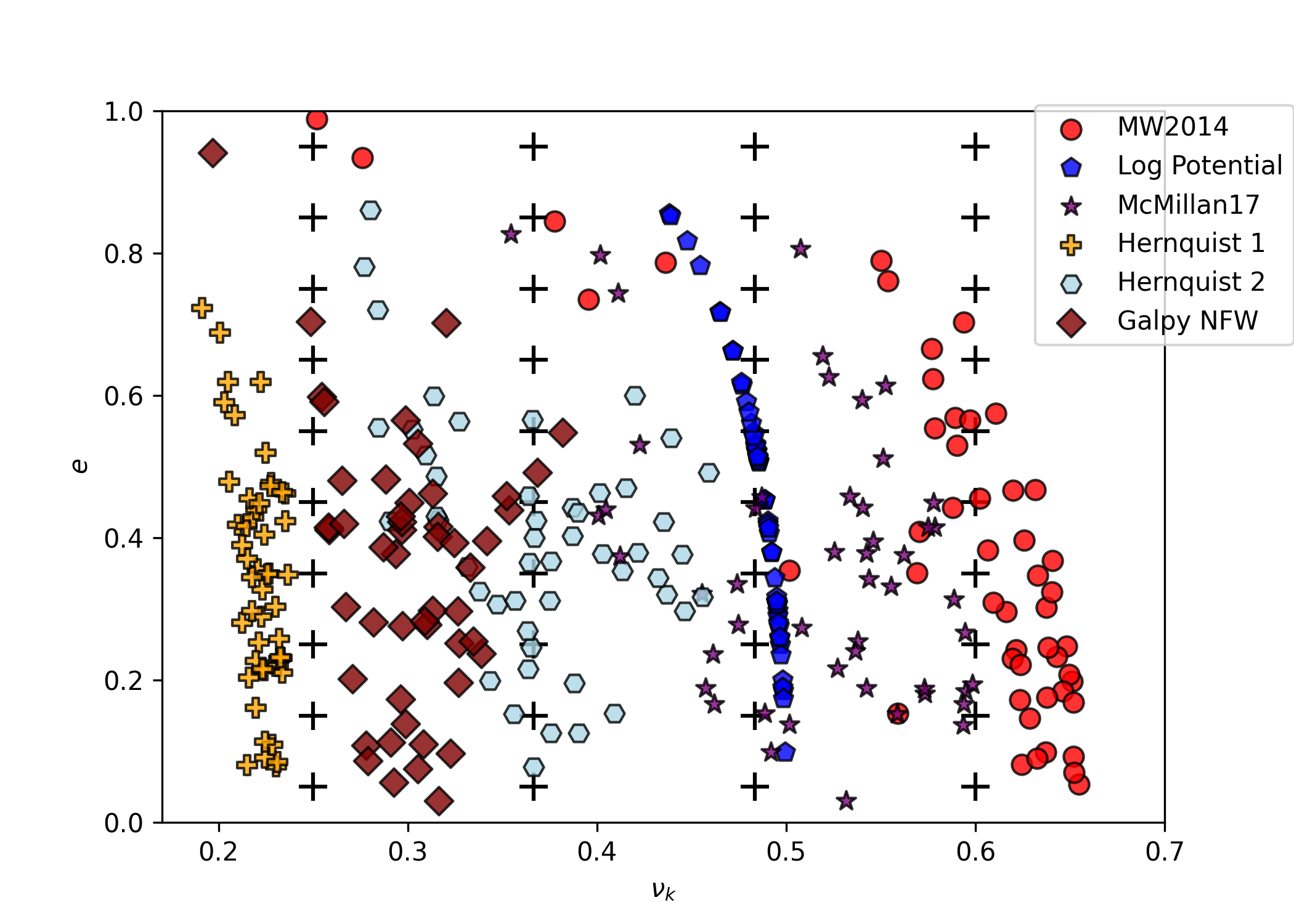}
     \caption{An illustration of the expansion used to estimate $t(\chi)$. The black `+' symbols represent values of $\nu_k$ and $e$ for which the $T_n$ integrals have been pre-computed. The colored symbols represent the $e$ and $\nu_k$ values of a set of randomly-selected orbits in 6 different potentials; the orbits are initialized at a galactocentric radius uniformly selected between 4 and 9 kpc, and their tangential velocity is selected uniformly from 0 to the $v_\mathrm{circ}$ value at that radius. The log potential (blue pentagons) has a fixed relationship between $e$ and $\nu_k$, but in general potentials will have some variation. As long as the grid of pre-computed $\nu_k$ and $e$ values has reasonable coverage for the orbits being considered, Equation \ref{eq:tau_n_chi_k_e} can be used to compute the $\tau_n$, which may in turn be used to compute $t(\chi)$ (Equation \ref{eq:tchi}).}
     \label{fig:enuk}
 \end{figure}
 
\begin{table}
    \centering
    \begin{tabular}{ccc}
       Name  & Parameters & Reference \\ \hline 
        MW2014 &  & \citet{bovy_Galpy_2015} \\
        Log Potential & $v_\mathrm{circ}=220\ \mathrm{km}\ \mathrm{s}^{-1}$ & \\
        McMillan17 & $R_0 = 8.21\ \mathrm{kpc}$, $v_0=233.1 \mathrm{km}\ \mathrm{s}^{-1}$ & \citet{mcmillan_Mass_2017}\\
        Hernquist 1 & $a=20\ \mathrm{kpc}$, $v_\mathrm{circ}(a)=200\ \mathrm{km}\ \mathrm{s}^{-1} $ & \citet{hernquist_Analytical_1990a}\\
        Hernquist 2 & $a=20\ \mathrm{kpc}$, $v_0=200\ \mathrm{km}\ \mathrm{s}^{-1}$ & \citet{hernquist_Analytical_1990a}\\
        Galpy NFW & $c=10$, $M_\mathrm{Vir}=10^{12}\ M_\odot$ & \citet{navarro_Universal_1997} \\
    \end{tabular}
    \caption{Parameters for the example potentials shown in Figure \ref{fig:enuk}. The parameters $R_0$ and $v_0$ refer to the galactocentric radius and circular velocity of the Sun (the adopted values of which affect Milky Way-specific potentials). Meanwhile $v_\mathrm{circ}$ is the circular velocity, which is constant for the Log Potential, and specified at the scale radius $a$ for the ``Hernquist 1'' potential. The NFW potential is set by its Virial mass $M_\mathrm{Vir}$ and concentration $c$, the ratio of the Virial radius to the scale radius.}
    \label{tab:potentials}
\end{table}

The precomputer therefore is responsible for computing $N_\mathrm{grid} \cdot N_e \cdot N_{\nu_k} \cdot (N_\mathrm{time}+2)$ integrals at $N_\chi$ points, where $N_\mathrm{grid}$ represents the number of points we use to span possible values of $e_0$ and $\nu_{k,0}$. Once evaluated the results of these numerical integrations can be applied for any particle in almost any potential, assuming that its value of $\nu_k$ and $e$ are not too distant from one of the pre-chosen $(\nu_{k,0},e_0)$ points. We demonstrate this graphically in Figure \ref{fig:enuk}, which shows a set of $N_\mathrm{grid}$ points in $(\nu_k,e)$ space where we have fixed $e_0$ and $\nu_{k,0}$ as black $+$ signs. The colored points show representative values of $\nu_k$ and $e$ for assorted orbits in 6 different potentials. The $N_\mathrm{grid}$ points cover enough of the parameter space that any orbit in these potentials can be safely evaluated to compute 
\begin{equation}
\label{eq:tchi}
    t(\chi) =  \frac{l^2}{h m_0 (1-e^2)^{\nu_k+(1/2)}} \left( 1+ \frac14 e^2 \left((w_0-w_2) \tau_0 +  (w_1-w_3)\tau_1 + \sum_{n=2}^{N_\mathrm{time}+2} (-w_{n-2}+2w_n-w_{n+2})\tau_n \right) \right),
\end{equation}
for $0\le\chi\le2\pi$, where we have suppressed the dependence of $\tau_n$ on $\chi$, $k$, $e$, $N_e$, and $N_{\nu_k}$ for brevity. We note that Lynden-Bell's version of this equation, in the text following his equation 4.37, uses an incorrect prefactor for the $\tau_1$ term.

Generally one wishes to know $\chi(t)$, not $t(\chi)$ so that $r$ and $\phi$ may be evaluated at a specified time. Inverting Equation \ref{eq:tchi}, a generalized version of Kepler's equation, can be done approximately \citep{lynden-bell_Approximate_2015}, but this approximation is not necessary for our purposes. We have enumerated $t(\chi)$ over a full radial oscillation $0\le \chi \le 2\pi$, so we can interpolate not only from $\chi$ to $t$, but just as easily, from $t$ to $\chi$, essentially solving the corresponding Kepler problem ``for free.'' We also note that each subsequent radial oscillation simply adds $T_r = t(\chi=2\pi)$ to $t$, so $t(\chi)$ in general is $N_\mathrm{orb} T_r + t(\chi - 2\pi N_\mathrm{orb})$, and $N_\mathrm{orb} = \mathrm{floor}(t/T_r)$. To evaluate Equation \ref{eq:phi}, it is also helpful to know the relationship between $\eta$ and $\chi$. By convention both are zero at pericenter. Lynden-Bell showed that
\begin{equation}
    1-e^2 = (1+e\cos\eta)(1-e\cos\chi).
\end{equation}
We can therefore find $\eta(\chi)$ with some care about matching signs by
\begin{equation}
\eta=
    \begin{cases}
        \arccos\left(\frac1e\left(\frac{1-e^2}{1-e\cos\chi} -1\right) \right) &\mathrm{if} \sin\chi\ge 0\\
        2\pi - \arccos\left(\frac1e\left(\frac{1-e^2}{1-e\cos\chi} -1\right) \right) &\mathrm{if} \sin\chi<0
    \end{cases}
\end{equation}
We can therefore start with a given $t$, then compute $\chi$ as above (which immediately tells us $r$ via Equation \ref{eq:U}), then $\eta$, which tells us $\phi$. Velocities can then immediately be computed via the conservation of angular momentum and energy.

\section{Vertical motion}
\label{sec:vertical}

Lynden-Bell's prescription works to arbitrary accuracy in potentials with spherical symmetry, namely $\psi(R,\phi,z)= \psi(R)$ (where $R=\sqrt{r^2+z^2}$ is the spherical radius). This is close to true for many practical cases (section \ref{sec:applications}), but for cases where the potential of a disk is important, we would like to generalize the $r-\phi$ motion we have computed so far to include motion in the $z$ direction. In this and the following sections, the $r-\phi$ and $x-y$ plane now refers to the midplane of the disk. 

In the textbook version of the epicyclic approximation \citep{ lacey_influence_1984, binney_Galactic_2008}, the equation of motion in the $z$ direction is
\begin{equation}
\label{eq:ddotz}
    \ddot{z} + \nu^2 z = 0,
\end{equation}
where $\ddot{z}$ denotes the second time derivative of $z$, and $\nu = \sqrt{4\pi G \rho_0}$ where $G$ is Newton's constant and $\rho_0$ is the midplane density of the disk. In the case where $\rho_0$ is constant, $\nu$ is constant and the $z$ motion is sinusoidal with frequency $\nu$ and an amplitude and phase set by the initial conditions. 

Generally $\nu$ will not, in fact, be constant, for two reasons. First, the quadratic form of the vertical potential is only correct when the vertical density profile is nearly constant \citep{binney_Galactic_2008}. The vertical extent of the particle's motion must therefore be much less than the disk scaleheight to neglect this issue reasonably. Second, the density of material in the disk is not constant with galactocentric radius, and the particle will be undergoing radial oscillations and therefore ``seeing'' different values of $\nu$. The second issue can be addressed by modifying the equation of motion and allowing $\nu$ to include a radial dependence. To address the first issue is more challenging because $\nu$ would need to depend on $z$ itself, making the equation nonlinear. The particles' trajectories become exactly periodic despite this nonlinearity if the change in $\nu$ with radius is neglected, so in this limit we could compute the particles' vertical motion by doing the numerical integration over a single half of the (initially unknown) vertical oscillation period.

We therefore have two mutually exclusive options. We can allow $\nu$ to vary with radius and require that the particle's height never far exceeds the scaleheight of the disk, or we can allow the particle to experience arbitrarily large vertical oscillations subject to an arbitrary vertical density profile, while not allowing this profile to change with galactocentric radius. The second option is problematic: any change that the particle experiences in its effective vertical oscillation frequency will break the symmetry required to propagate the particle many vertical oscillations into the future, with substantial errors accumulating from one vertical oscillation to the next. This method also requires numerical integrations. We therefore focus on the first option.

In this case, $\nu(t)$ is periodic with period $T_r$. This makes Equation \ref{eq:ddotz} a special case of Hill's equation \citep{hill_Part_1877}. These equations may be solved in general by finding two independent solutions (typically numerically) over a single (in this case radial) period. This method is underpinned by Floquet Theory \citep[e.g.][]{magnus_Hill_1966} which guarantees that the general solution may be written
\begin{equation}
\label{eq:genz}
    z = A e^{\mu t} f(t) + B e^{-\mu t} f(-t),
\end{equation}
where $f(t)$ is a function with the same period as $\nu(t)$, and $A$ and $B$ are constants to be set by the initial conditions ($z$ and $v_z$). WKB approximations, as discussed for this problem by the ``B'' of WKB \citep{brillouin_Practical_1948} and as the basis of numerical integrators \citep{agocs_Adaptive_2022} may also be useful, though they tend to work best when there are many oscillations of $z$ for each radial oscillation, i.e. $\nu \gg \kappa$, which is not quite true in the Solar neighborhood. Another method of solution proposed by \citet{fiore_Timedependent_2022} involves writing down a Volterra series of integrals for the phase of the $z$-oscillation. Hill's original method involves decomposing $\nu(t)$ into a Fourier series from which $\mu$ can be computed, and $f(t)$ can be found as a related Fourier series. We implement the following three methods in \lbparticles{}:
\begin{itemize}
\item ``2Int'' -- numerically integrate the equation of motion for $z$ over a single radial period.
\item ``Fourier'' -- decompose $\nu(t)$ into a Fourier series, from which $\mu$ and $f(t)$ may be estimated.
\item ``Volterra'' -- use the prescription of \citet{fiore_Timedependent_2022} to estimate $z$, based on a series of integrals of the phase of the oscillation.
\end{itemize}
In each case we of course need to specify $\nu^2(r)$. Physically $\nu^2 = 4\pi G \rho_0$, and $\rho_0 \sim \Sigma_*/H_*$, where $\Sigma_*$ is the surface density of stars in the disk at a given galactocentric radius, and $H_*$ is the scaleheight of the stars. Empirically many disk galaxies have $H_* \sim \mathrm{const}$ with radius \citep[e.g.][]{vanderkruit_Surface_1982,tsukui_Emergence_2024}, so we might expect $\nu^2$ to scale the same way as $\Sigma_*$, which could reasonably locally be represented by an exponential or a powerlaw \citep[e.g.][]{forbes_balance_2014}. We now discuss each method in more detail.

\subsection{The 2Int Method}

Here we follow the textbook presentation in chapter 3.6 of \citet{teschl_Ordinary_2012}. They define the principal matrix solution of Equation \ref{eq:ddotz} as
\begin{equation}
\Pi(t,t_0) = 
    \begin{bmatrix}
c(t,t_0) & s(t,t_0)\\
\dot{c}(t,t_0) & \dot{s}(t,t_0) \\
\end{bmatrix}.
\end{equation}
The functions $s$ and $c$ are solutions to the initial value problem as a function of time $t$ where $t_0$ denotes the value of $t$ at which the initial conditions are specified. The two solutions $s$ and $c$ are defined such that $\Pi(t_0,t_0)$ is the identity matrix. That is, at $t_0$, $c=1$, $s=0$, $\dot{c}=0$, and $\dot{s}=1$. The monodromy matrix is then defined as $M(t_0) = \Pi(t+T_r,t_0)$. The monodromy matrix and its characteristic equation are crucial in evaluating the stability of the solutions. For our problem, $\mu$ is always purely imaginary, so each of the two independent solutions advances by a particular phase dependent on the orbit, but does not grow in overall amplitude. Once $c$, $s$, $\dot{c}$, and $\dot{s}$ are tabulated from $0\le t\le T_r$, we can immediately obtain $z$ and $\dot{z}$ at any time in the future via
\begin{equation}
    \begin{bmatrix}
        z(t)\\
        \dot{z}(t)
    \end{bmatrix}
    = \Pi(t - N_\mathrm{orb} T_r, t_0) \cdot \left(M(t_0) \right)^{N_\mathrm{orb}} \cdot 
    \begin{bmatrix}
    z(t_0) \\
    \dot{z}(t_0)
    \end{bmatrix}.
\end{equation}
This method is straightforward and reliable, but relies on doing (two) numerical integrations to find $c$ and $s$. The integrations need only be carried out over a single radial period, which for stars in the Solar neighborhood is around 160 Myr, and only requires the integration of a pair of coupled ODEs (for $z$ and $\dot{z}$) rather than 6 for the full trajectory of the particle. 

\subsection{The Fourier Method}

Hill's original method \citep{hill_Part_1877} involved decomposing $\nu(t)$ into its Fourier components,
\begin{equation}
\label{eq:nuexpansion}
    \nu^2 \approx \theta_0 + 2\sum_{n=1}^{N_{z,\mathrm{fourier}}} \theta_n \cos( n\ 2\pi t/T_r),
\end{equation}
which can be estimated in nearly the same way as $W$ (see Equations \ref{eq:Wseries} and \ref{eq:Cmatrix}). Note the following two conventions in early 20th-century work on Hill's equation: first, that the period of the oscillation is $\pi$ rather than $2\pi$, and second, that the $\theta_i$ are a factor of 2 smaller than the more-familiar Fourier expansion.

Following \citet{whittaker_Course_1927}, we then assume that one solution to the equation is given by
\begin{equation}
    u = e^{\mu t} \sum_{n=-\infty}^\infty b_n e^{2n i t/T_r},
\end{equation}
we can plug this and Equation \ref{eq:nuexpansion} into the original differential equation and match terms of the same order to find the following infinite system of homogeneous linear equations
\begin{equation}
\label{eq:bn}
    (\mu + 2n i)^2 b_n + \sum_{m=-\infty}^\infty \theta_m b_{n-m} = 0
\end{equation}
for any integer value of $n$. Note that $\theta_{-n} = \theta_n$, and we have taken $\theta_n = 0$ for any $n>N_{z,\mathrm{fourier}}$.

In order for this system of equations to have (a family of) solutions besides the trivial solution ($b_n=0$ for all $n$), the determinant of the matrix defining this infinite system must be zero. This places the following condition on $\mu$:
\begin{equation}
    -\sinh^2(\mu \pi/2) = \Delta_1(0) \sin^2(\sqrt{\theta_0}\pi/2),
\end{equation}
where $\Delta_1(0)$ is the determinant of the matrix with the following components
\begin{equation}
    B_{mp} = 
    \begin{cases}
        \frac{\theta_{m-p}}{\theta_0 - 4 m^2}\ &\mathrm{m \ne p} \\
        1 & m=p
    \end{cases}.
\end{equation}
In principle this matrix is infinite even if our Fourier series for $\nu^2$ is truncated at a finite order, but in practice the determinant can be evaluated for a finite version of $B_{mp}$ that may only be $\sim$ twice as large in each dimension as $N_{z,\mathrm{fourier}}$.

Once $\mu$ is estimated, we can numerically evaluate each component of the design matrix of Equation \ref{eq:bn}. Again, this is in principle an infinite matrix even if the series for $\nu^2$ is not infinite, but high-order terms are suppressed by $\sim n^2$, when $|n|$ is large. In principle our choice of $\mu$ is such that the determinant of the design matrix is zero. Numerically this is not exactly true, but the design matrix can be expanded in a singular value decomposition, which will identify the small singular value and a corresponding non-trivial solution family to Equation \ref{eq:bn}, i.e. a set of $b_n$ that may be scaled by any real number and still yield zero on the right-hand side. These $b_n$ then define the periodic component of the solution, $\mu$ defines the exponential component, and all that remains is to solve the pair of equations given by Equation \ref{eq:genz} and its first derivative for $A$ and $B$ given initial values of $z$ and $\dot{z}$.

\subsection{The Volterra Method}
\label{sec:volterra}
\citet{fiore_Timedependent_2022} points out that the second-order ODE for $\ddot{z}$ can be transformed into two first-order equations:
\begin{eqnarray}
\label{eq:psidot}
    \dot{\psi} &=& \nu + \frac{\dot{\nu}}{2\nu} \sin(2\psi) \\
    \frac{\dot{\mathcal{I}}}{\mathcal{I}} &=& -\frac{\dot{\nu}}{\nu} \cos(2\psi)
\end{eqnarray}
where critically the first can be solved without reference to the second. Here $\psi$ refers to the phase of the oscillation, and $\sqrt{\mathcal{I}}$ is proportional to the amplitude (more specifically $\mathcal{I}=H/\nu$, where $H=(\dot{z}^2 + \nu^2(t) z^2)/2$ is the Hamiltonian). This transformation reduces the problem essentially to a single first-order ODE for $\psi$, but the ODE is non-linear, with $\psi$ advancing both via the time-dependent $\nu$ and an oscillating non-linear term $\dot{\nu}\sin(2\psi)/(2\nu)$.

\citet{fiore_Timedependent_2022} provides a series solution to this pair of ODEs. The 0th order approximation is given by
\begin{eqnarray}
    \mathcal{I}^{(0)} &=& \mathcal{I}(t_0) \\
    \psi^{(0)} & \equiv & \varphi = \psi(t_0) + \int_{t_0}^t dt' \nu(t') \\
    \label{eq:z0}
    z^{(0)} &=& \sqrt{\frac{2\mathcal{I}^{(0)}}{\nu}} \sin(\psi^{(0)}) \\
     \dot{z}^{(0)} &=& \sqrt{2\mathcal{I}^{(0)}\nu} \cos(\psi^{(0)})
\end{eqnarray}
The values of $\psi(t_0)$ and $\mathcal{I}(t_0)$ can be computed directly respectively by
\begin{equation}
    \psi = \mathrm{arctan2}(\dot{z}, \nu z)
\end{equation}
and by plugging the initial conditions into the definition of $\mathcal{I}$ respectively.

Higher-order corrections can be computed using a Volterra-type equation
\begin{equation}
\label{eq:psihigh}
    \psi^{(i)}(t) = \varphi(t) + \int_{t_0}^t \frac{\dot{\nu}(t')}{2\nu(t')}\sin(2\psi^{(i-1)}(t')) dt',
\end{equation}
from which corrections to $\mathcal{I}$ can also be computed:
\begin{equation}
    \mathcal{I}^{(i)}(t) = \mathcal{I}(t_0) \exp\left(-\int_{t_0}^t \frac{\dot{\nu}(t')}{\nu(t')} \cos(2\psi^{(i-1)}(t'))dt'\right).
\end{equation}
Essentially each iteration of $\psi$ is passed back in to Equation \ref{eq:psihigh}, which provides a better accounting of the effects of the oscillatory term in Equation \ref{eq:psidot}. The 0th order approximation to $\psi$ neglects the oscillatory term, but accounts for the main (non-oscillatory) term contributing to $\dot{\psi}$ and the effects of the varying value of $\nu$ on the amplitudes of $z$ and $\dot{z}$.

The question then becomes whether it is feasible to compute any of these integrals, starting with $\int \nu(t) dt$, in a way that is competitive with the other two methods already discussed. With a suitable choice for the form of $\nu(r)$, it is possible to evaluate $\varphi$, the 0th order phase, with an integral almost identical to the one discussed in Section \ref{sec:t}. If we assume that $\nu$ is a simple powerlaw with respect to $r$, then the only new thing in the integrand, $\nu$, simply becomes an additional factor of $(1-e\cos\chi')$ raised to a power related to the powerlaw index of $\nu$ with $r$. Noting the similarity to Equation \ref{eq:tint}, we have
\begin{equation}
\label{eq:varphi}
    \varphi = \nu_0 \left(\frac{r_0}{\bar{U}}\right)^{\alpha/(2k)} \frac{l^2 }{h m_0 (1-e^2)^{\nu_k+(1/2)}} \int_0^\chi (1-e\cos\chi')^{\nu_k-\alpha/(2k)} (1+w) d\chi',
\end{equation}
where $\nu_0$, $r_0$, and $\alpha$ are all defined via
\begin{equation}
    \nu = \nu_0 \left(\frac{r}{r_0} \right)^{-\alpha/2}.
\end{equation}
If we know $\alpha$, we can therefore pre-compute $\phi(\chi)$ in exactly the same way we did for $t(\chi)$. When constructing the precomputer, we therefore calculate not just the integrals defined in Equation \ref{eq:integrands}, but a very similar set
\begin{equation}
    \label{eq:pj}
    \mathcal{P}_n(\chi, k,e;j,m) = \int_0^\chi (1 - e_0 \cos\chi')^{\nu_k-j-\alpha/(2k)}(\cos\chi')^j\cos(n\chi') B^m d\chi',
\end{equation}
and reconstruct $\varphi(\chi)$ using the same set of coefficients $w$ as used in $t(\chi)$. The only difference is that the 2D Taylor expansion is done with respect to $\mu_k = \nu_k-\alpha/(2k)$ instead of $\nu_k$, so the integral includes powers of
\begin{equation}
    B = \left( \psi^{(0)}(\mu_{k_0}+1) - \psi^{(0)}(\mu_{k_0}-j+1) + \ln(1-e_0\cos\chi)\right)
\end{equation}
rather than $A$ as defined in Equation \ref{eq:Adef}.

As with $t(\chi)$, this pre-computation approach allows us to use the same set of pre-computed integrals for a particle with any orbit, provided the set of pre-computed $\mu_{k_0}$ and $e_0$ has been chosen such that $\delta e = e-e_0$ and $\delta\mu_k=\mu_k-\mu_{k_0}$ are not excessively large.

The higher-order corrections, however, cannot be precomputed because the prefactor in Equation \ref{eq:varphi} needs to be included in the integrand (Equation \ref{eq:psihigh}), so the integrals needs to be computed for each new particle. As with the 2Int method, the integral need only be computed for $0 < \chi \le 2\pi$. Each subsequent order requires another factor of 2 more integrals and a corresponding quantity of tedious algebra (see Appendix \ref{sec:volterra1st}), so this method is not suitable for applications that require high precision and hence higher orders of $\psi$.

\section{Results and benchmarking} 
\label{sec:results}

To use Lynden-Bell's approach in practice it will be helpful to know the error in position of an \lbparticles{} solution. To evaluate this, we must first write down the equations of motion and perform an integration with restrictive tolerances to establish a ``ground truth'' against which to compare.
\begin{align}
    \ddot{x} &= \frac{d \psi}{dr} \frac{x}{r} \\
    \ddot{y} &= \frac{d \psi}{dr} \frac{y}{r} \nonumber \\
    \ddot{z} &= -\nu^2 z \nonumber
\end{align}
These second-order equations in time can be transformed to first-order equations in the usual way by introducing $\dot{x}$, $\dot{y}$, and $\dot{z}$ as separate variables whose derivatives are given above, and which themselves are the derivatives of $x$, $y$, and $z$. Note that the radial derivatives of $\psi$ are negative because of the sign convention of $\psi$, and that the $r$ appearing in the denominator of the first two lines is the in-plane or cylindrical radius only, since this is the set of equations being solved by the \lbparticles{}. If instead the equations of motion use the spherical radius and a component of the vertical acceleration from the background spherical potential, \lbparticles{} can handle this case approximately with a small modification to the potential (see Appendix \ref{sec:nonsep}).

 \begin{figure}
     \centering
     \includegraphics[width=\textwidth]{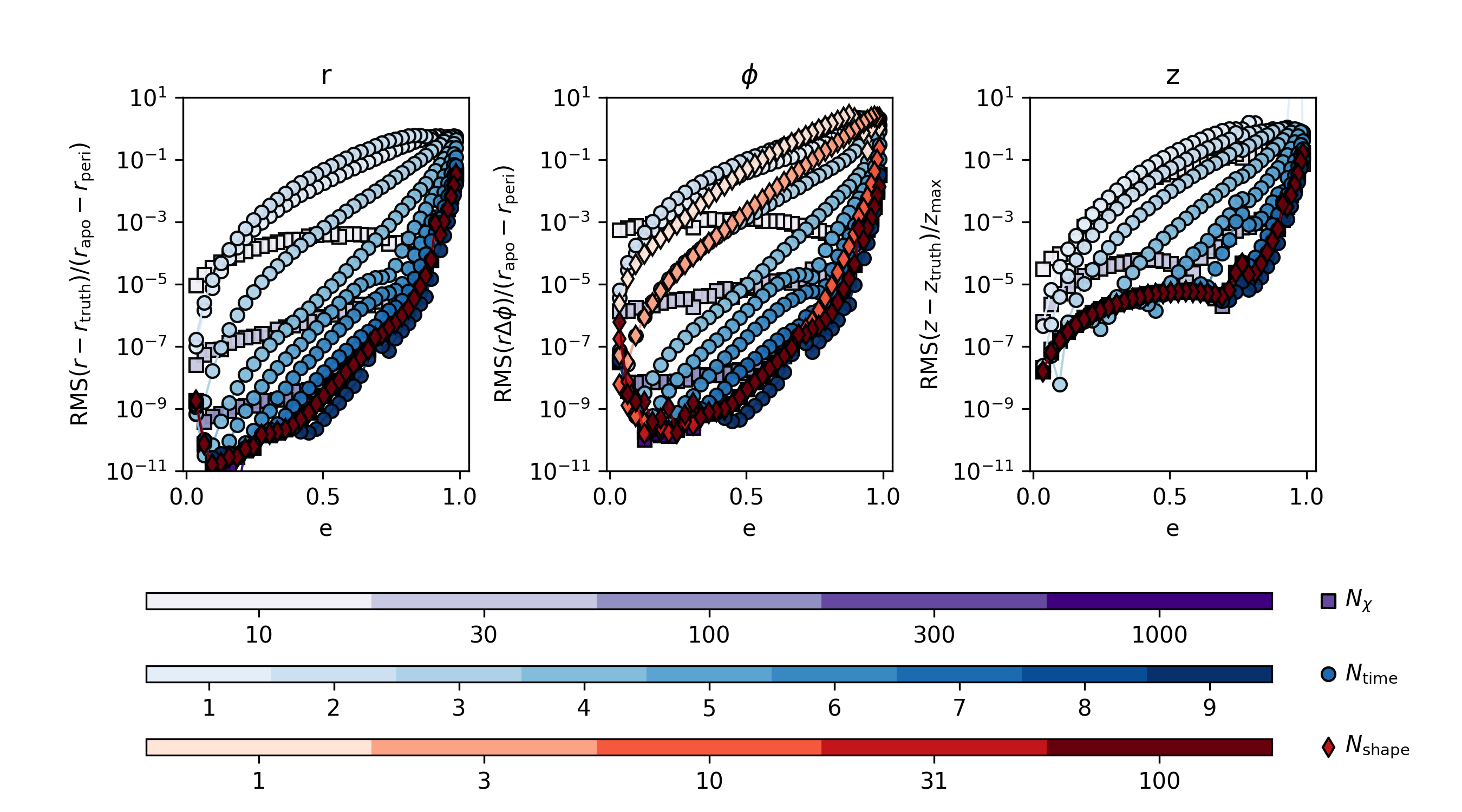}
     \caption{Errors in the $r$, $\phi$, and $z$ components of \lbparticles{} solutions relative to a high-precision numerical integration of the same orbits. Purple squares show variation in $N_\chi$, blue circles variation in $N_\mathrm{time}$, and red diamonds in $N_\mathrm{shape}$. When they are not being varied (i.e. when one of the other two parameters is being varied), these parameters are set to $N_\chi=300$, $N_\mathrm{time}=8$, and $N_\mathrm{shape}=15$.}
     \label{fig:errors}
 \end{figure}

We evaluate errors by comparing the ground truth integration to the \lbparticles{} solution on a grid of evaluation times. The grid of times is equally spaced in 10 Myr intervals from 0 to 10 Gyr. At each point, we evaluate the errors in the cylindrical components of the particle's position as $\delta r = r - r_\mathrm{truth}$, $\delta\phi_\mathrm{simple} = \phi - \phi_\mathrm{truth}$, and $\delta z = z - z_\mathrm{truth}$. Because $\phi$ is periodic from $0$ to $2\pi$, we replace values of $\delta\phi_\mathrm{simple}$ greater than $\pi$ with $2\pi-\delta\phi_\mathrm{simple}$, so that
\begin{equation}
    \Delta\phi = \begin{cases}
        \delta\phi_\mathrm{simple}\ (\mathrm{mod}\ 2\pi)\ &\mathrm{if}\ \delta\phi_\mathrm{simple} \ (\mathrm{mod}\ 2\pi) \le \pi \\
        2\pi - (\delta\phi_\mathrm{simple}\ (\mathrm{mod}\ 2\pi))\ &\mathrm{if}\ \delta\phi_\mathrm{simple} \ (\mathrm{mod}\ 2\pi) > \pi. \\
    \end{cases}
\end{equation}
Each quantity is normalized by the largest error we might expect in that direction: $r_\mathrm{apo}-r_\mathrm{peri}$ in the $r$-direction, $(r_\mathrm{apo}-r_\mathrm{peri})/r$ in the $\phi$-direction, and $z_\mathrm{max}$, the maximum absolute value of $z_\mathrm{truth}$ over the integration. Finally, we take the root mean square of the normalized quantities over the 100 time points in the final Gyr of the simulation. We plot these errors for a grid of orbits in the MW2014 \citep{bovy_Galpy_2015} potential as a function of their eccentricity $e$. The orbits are initialized at $t=0$ with $(x,y,z)=(8100,0,21)$ pc, and $(v_x,v_y,v_z)=(0,v_\phi,10)\ \mathrm{pc}\ \mathrm{Myr}^{-1}$, with $v_\phi$ equally spaced between 19 and 219 $\mathrm{pc}\ \mathrm{Myr}^{-1}$. This produces eccentricities from $0.035$ to $0.986$. For the $z$ evolution of the particles, we use the Fourier method with $N_{z,\mathrm{fourier}}=80$. The results are shown in Figure \ref{fig:errors}.

We vary three numerical parameters to assess their effect on the errors. The first is $N_\chi$, the number of values of $\chi$ between $0$ and $2\pi$ used to generate the interpolation between $t$ and the particle's phase in its transformed radial oscillation, $\chi$. We also vary $N_\mathrm{time}$ and $N_\mathrm{shape}$, respectively the number of terms retained in the expansions corresponding to $t$ and $\phi$. We find that increasing $N_\chi$ has diminishing returns for values above $\sim 100$, while adding more terms to the $t$ expansion, i.e. increasing $N_\mathrm{time}$, systematically improves performance across all 3 variables and all values of $e$ -- this is because all 3 variables rely on the mapping between $t$ and $\chi$ to evaluate the particle's position. Meanwhile $N_\mathrm{shape}$ only matters for the error in $\phi$, and experiences diminishing returns above values of $\sim 10$. At the highest eccentricities, the errors can reach up to 10\%, as might be expected for an expansion based on the epicyclic approximation, while for more typical disk-like orbits, the errors are exceptionally small. The upturn in the dimensionless error values at small values of $e$ is because $r_\mathrm{apo}-r_\mathrm{peri}$ approaches 0 as $e$ approaches 0.

\section{Discussion}
\label{sec:discussion}
\label{sec:applications}

Lynden-Bell's approach to writing down orbits and our implementation of it are generally useful because conserved quantities of the orbit remain conserved by construction. This is not guaranteed in generic integrations of the equations of motion. Moreover the cost to evaluate a particle's position and velocity at some time $t$ arbitrarily far into the future (under the simplifying assumption that the potential is constant in time) is independent of $t$, whereas generic integrations require a runtime proportional to $t$. Depending on parameter choices, an \lbparticles{} object may be a more compact representation of an orbit, since a direct integration requires storing enough time points that the particle's location at a particular time may be reconstructed via interpolation. Finally, the quantities computed as part of the implementation, including $r_\mathrm{peri}$, $r_\mathrm{apo}$, $e$, $k$, $m$, and the $W_i$, and $w_i$, may be useful as features to understand families of orbits. We now mention a few interesting potential use cases, and the limitations of this approach.

\subsection{Applications}

\subsubsection{Interstellar Object Orbits}
The nascent field of interstellar objects (ISOs) is a direct application of this formalism. 
Interstellar objects are planetesimals that originate from stellar systems of sufficient metallicity, throughout Galactic history \citep{ISSITeam:2019}, and are therefore sensitive tracers of Galactic stellar formation history \citep{Lintott_2022} and dynamics.
Planetary formation and evolution processes mean that systems are efficient donors to the ISM, e.g. in the Solar System planetary architecture, 75-85\% of the initial cometary bodies are now unbound \citep{Fernandez_1984, Brasser_2006}.
As unbound planetesimals should outlive their star, the Galactic ISO population is formed from the integrated population contributed from all stellar lifetimes, modulo metallicity: the \textit{sine morte} population \citep{Hopkins:2023, hopkins_Predicting_2025}.
The propagation of ISOs from each of their progenitor systems will then occur in the gravitational potential of the Galaxy, and follow the motion that we describe here.
In \citet{forbes_He_2024} we employ \lbparticles{} to understand the density structure of streams of interstellar objects. 

\subsubsection{Stellar Streams}

While streams in the disk are subject to heating and the difficulties with analytically modelling the vertical motion of the particles, most observed stellar streams orbit far from the disk, where a lack of heating allows them to maintain low velocity dispersions and hence high coherence \citep[see][for a review]{bonaca_Stellar_2024}. Streams, as a population of similar orbits, are good candidates for analytic, or at least semi-analytic, descriptions, for use in generative models \citep{bovy_Dynamical_2014}, and understanding their morphologies \citep{kupper_Structure_2008, amorisco_Feathers_2015}. This is of broad interest because structures in the streams may correspond to strong interactions with individual massive objects including dark matter substructure \citep{bovy_Detecting_2016,bonaca_Spur_2019}. Streams may also be detected in other galaxies \citep[e.g.][]{nibauer_Constraining_2023}. However, streams near resonance with the bar \citep{pearson_Gaps_2017}, and that are sensitive to asymmetric features in the potential in general, e.g. triaxiality \citep[e.g.][]{law_Evidence_2009} require a more flexible description of the orbits.

\subsubsection{Dark Matter orbits}

The orbits of dark matter particles themselves may also be understood with a simple orbital prescription. The orbits of these particles will typically be only minimally influenced by the disk, given the halo's much greater extent. In \citet{dsouza_Enhanced_2024}, we used \lbparticles{} to enumerate the orbits of axion mini-halos, objects that form under certain circumstances if the QCD axion represents the bulk of the dark matter in the Universe. These mini-halos may be disrupted by their interactions with the stellar disk \citep{shen_Disruption_2024}, which in turn may produce tidal streams \citep{ohare_Axion_2023}, which in turn affects the probability that direct detection experiments on Earth are actually sensitive to this type of dark matter.

Even if the dark matter particles are not organized within mini-halos, their individual orbits are of interest when attempting to understand the dynamics of any sort of cold dark matter. The apocenters of orbits, and the identification of related orbits are used by the SPARTA code \citep{diemer_Splashback_2017,diemer_Splashback_2020} to identify the Splashback radius of halos, and track subhalos respectively. A semi-analytic orbit description may aid in these and related endeavours once on-the-fly halo identification has proceeded, and a corresponding approximately spherical potential derived.

\subsection{Cautions}
\label{sec:cautions}
The \lbparticles{} package solves the following problem: the motion of a test particle under the gravitational influence of a (reasonably) smooth time-invariant potential. The potential consists of a spherically-symmetric component, and may optionally include an axisymmetric disk-like component. Motion resulting from the disk component can only be modeled if the vertical oscillation frequency $\nu$ can be expressed purely as a function of the cylindrical galactocentric radius $r$. For a self-gravitating disk, this condition is only true when $z/H \ll 1$, where $H \equiv \Sigma/(2\rho_0)$ is the scaleheight \citep[see][for instance]{hassan_Implementation_2024}, $\Sigma$ is the surface density of the disk and $\rho_0$ is its midplane density. 

We also assume that the particle's vertical position does not affect its radial motion, which, again is only exactly true when the particle remains close to the midplane. Corrections for the spherical (as opposed to cylindrical) nature of the potential can be computed (see Appendix \ref{sec:nonsep}), but these corrections are just averages, and the particle's radial motion takes place in slightly different effective potentials for each orbit. For both of these reasons, caution is warranted for particles whose motion takes them well above the Galactic midplane.

Yet another practical hazard of computing orbits in the Galactic disk is the presence of dynamical heating, wherein gravitational interactions between the particle and massive structures in the disk tend to cause the particles to increase both their vertical energy and the component of their energy associated with in-plane but non-circular motions (i.e. the difference between the star's in-plane velocity and the circular velocity). Dynamical heating is sourced by non-axisymmetric non-constant components in the potential, so clearly this motion is not covered under the ``warranty.'' It may be possible to model this effect by allowing the particle's properties to random walk over time as appropriate, but it is not obvious how all of the properties we compute in \lbparticles{} evolve under this heating. An action-angle diffusion approach \citep[see e.g.][]{arunima_Their_2025} may be more promising.

\section{Conclusion} \label{sec:conclusion}

In this work we have taken a practical step forward in making the orbits of individual particles in a galactic potential more legible. Orbits can be quickly and compactly described in a numerical form to high accuracy for any spherically-symmetric potential. This work corrects errors in \citet{lynden-bell_Bound_2015}'s work that may have made implementing his results difficult, and includes a step forward in being able to jointly model the in-plane and vertical motion of particles embedded in a self-gravitating disk. We also include a practical numerical mapping between $t$ and the phase of the transformed radial oscillation, $\chi$, in order to solve the analogous Kepler equation. The code is freely available and pip installable as \lbparticles{}. \\

M.T.B. and J.C.F. appreciate support by the Rutherford Discovery Fellowships from New Zealand Government funding, administered by the Royal Society Te Ap\={a}rangi. 
J.C.F. thanks Kathryn Johnston, David Spergel, and Matthew Holman for helpful discussions.

\software{numpy \citep{vanderwalt_numpy_2011,harris_array_2020},
          matplotlib \citep{hunter_matplotlib_2007},
          scipy \citep{virtanen_scipy_2020},
          }

\appendix

\section{Corrections for Non-Separable In-Plane and Vertical Motion}
\label{sec:nonsep}
Throughout the text we made the simplifying assumption that the in-plane motion of the particle was not influenced by the particle's $z$ position. For a particle orbiting under the gravitational influence of both a spherically-symmetric matter distribution and a disk, there are two differences in the equations of motion:
\begin{align}
    \ddot{x} &= \frac{d \psi}{dR}(R) \frac{x}{R}, \\ \nonumber
    \ddot{y} &= \frac{d \psi}{dR}(R) \frac{y}{R}, \\ \nonumber
    \ddot{z} &= \frac{d \psi}{dR}(R) \frac{z}{R} -\nu^2 z,
\end{align}
where $R$ is the spherical radius $\sqrt{x^2+y^2+z^2}$. First, each equation uses $R$ rather than the cylindrical radius $r=\sqrt{x^2+y^2}$, and second, the $z$-component of the acceleration includes the effect of the spherical potential (the first term), in addition to the local gravitational force of the disk encapsulated in the second term $\propto \nu^2$. To leading order when $z \ll r$ these sets of equations of motion are identical to the ones used in Section \ref{sec:results}, but here we derive the next-order correction. 

For the in-plane components of the acceleration, we have for instance
\begin{equation}
    \frac{d\psi}{dR}(R)\ \frac{x}{R} = \left(\frac{d\psi}{dr}(r) + (R-r)\frac{d^2\psi}{d r^2}(r) + \cdots \right) \frac{x}{r} \left( 1 - \frac12 \left(\frac{z}{r}\right)^2 + \frac38\left(\frac{z}{r}\right)^4 + \cdots \right)
\end{equation}
where
\begin{equation}
    R-r = \sqrt{r^2+z^2} - r = r \sqrt{1+(z/r)^2} - r = r((1/2)(z/r)^2 - (1/8)(z/r)^4 + \cdots).
\end{equation}
and
\begin{equation}
    (R-r)^2 = (\sqrt{r^2+z^2} - r)^2 = r^2 (\sqrt{1+(z/r)^2} - 1)^2 = r^2 ( (1+(z/r)^2) - 2\sqrt{1+(z/r)^2} + 1 ) = r^2 ( (z/r)^2 - 2((1/2)(z/r)^2+\cdots) )
\end{equation}

The in-plane accelerations may be corrected by adding the following to the cylindrical version of the acceleration $(d\psi/dr) (x/r)$,
\begin{equation}
    \Delta a_x =  \frac{x z^2}{2r^3} \left( r \frac{d^2 \psi}{dr^2} -  \frac{d\psi}{dr} \right),
\end{equation}
for the $x$-component, and similarly for the $y$-component. We can therefore define a correction to $\psi$ which we will call $\delta \psi$, via
\begin{equation}
\label{eq:deltapsi}
    \frac{d\delta\psi}{dr} = \frac{z^2}{2r^2} \left( r \frac{d^2 \psi}{dr^2} -  \frac{d\psi}{dr} \right).
\end{equation}
The particles will then move in the corrected potential $\psi + \delta\psi$.

The particle's $z$-position is of course time-dependent, but assuming that $\nu/\kappa$ is irrational, the typical $z^2$ can be estimated based on the particle's vertical energy and the $r$-dependent $\nu$. The first and second derivatives of the unmodified potential are already known since they are used to efficiently find the peri- and apo-center of the particle's orbit, so to include this modification to the potential we need to compute $d^3\psi/dr^3$ and the integral of the right hand side of Equation \ref{eq:deltapsi}.

For the vertical acceleration, the first-order correction can be absorbed into $\nu^2$,
\begin{equation}
    \frac{d\psi}{dR}(R) \frac{z}{R} - \nu^2 z \approx -\left(\nu^2 - \frac{z^2}{2r^3} \left( r \frac{d^2 \psi}{dr^2} -  \frac{d\psi}{dr} \right) \right) z,
\end{equation}
where as with the in-plane component we approximate $z^2$ by its radially-dependent time-average.

To estimate the average value of $z^2$ at each radius, we make use of the 0th order solution for $z$ of section \ref{sec:volterra} from Equation \ref{eq:z0}. If we assume that $\psi$ is uniformly distributed, the average value of $z^2$ will be
\begin{equation}
    \langle z^2 \rangle(r) \approx \frac{\mathcal{I}(t_0)}{\nu(r)},
\end{equation}
where all we have done is approximate $\langle \sin^2\varphi \rangle \approx 0.5$

We need to know the value and first two derivatives of the perturbed potential $\psi + \delta\psi$.
\begin{equation}
    \delta \psi \approx \mathcal{I}(t_0) \int_{r_\mathrm{min}}^r \frac{1}{2r'^2 \nu(r')} \left(r'\frac{d^2\psi}{dr'^2} - \frac{d\psi}{dr'} \right) dr'
\end{equation}
This integral is computed once (numerically) for each combination of $\nu$ and $\psi$. The first derivative is just a matter of substituting in our approximate value of $z^2$ to Equation \ref{eq:deltapsi}:
\begin{equation}
        \frac{d\delta\psi}{dr} = \frac{\mathcal{I}(t_0)}{2\nu(r)r^2} \left( r \frac{d^2 \psi}{dr^2} -  \frac{d\psi}{dr} \right).
\end{equation}
Finally, the second derivative requires a little algebra to write in terms of a general $\nu$, and will require us to know both $d^3\psi/dr^3$ and $d\nu/dr$. The second derivative of the perturbation is
\begin{equation}
    \frac{d^2\delta\psi}{dr^2} =\mathcal{I}(t_0) \frac{1}{2\nu r^2} \left[ \left( -\frac{d\ln\nu}{dr} - \frac{2}{r} \right)\left( r \frac{d^2 \psi}{dr^2} -  \frac{d\psi}{dr} \right)  +  r\frac{d^3 \psi}{dr^3} \right].
\end{equation}

\section{First-Order Correction for Volterra Method}
\label{sec:volterra1st}

The higher-order corrections for the Volterra method require $2^N$ integrations for each particle where $N$ is the order, so it is not worth using higher orders. We have nonetheless implemented the first-order correction to demonstrate how the higher orders would be computed. The integrals that we need to compute are of the form
\begin{equation}
\label{eq:1storder}
    I_T \equiv \int_{t_0}^t \frac{\dot{\nu}(t')}{\nu(t')} T(2\varphi(t')) dt',
\end{equation}
where $T$ is either cosine or sine (the sine integrals are necessary to compute the higher-order phases, and the cosine integrals are necessary for the corrections to the amplitude). For now we will focus on $T(x)=\sin(x)$. While we expect some numerical integration will be necessary, we want to restrict it to a single radial period. We therefore split the 0th order phase $\varphi$ into terms that vary in time and those that are constant.
\begin{equation}
    \varphi(t) = \varphi(t_0) + \int_0^{\chi(t)} f(\chi') d\chi' - \int_0^{\chi(t_0)} f(\chi') d\chi' + N_\mathrm{orb}\int_0^{2\pi} f(\chi')d\chi'.
\end{equation}
Here $f(\chi) = \nu dt/d\chi$. Within a given radial period, $N_\mathrm{orb}$ is constant. The first, third and fourth terms are all therefore constants. We abbreviate these constant terms
\begin{equation}
    \varphi_\mathrm{IC} = \varphi(t_0) - \int_0^{\chi(t_0)} f(\chi') d\chi',
\end{equation}
the phase associated with the initial conditions, and 
\begin{equation}
    \varphi_{T_r} = \int_0^{2\pi} f(\chi')d\chi,
\end{equation}
the total 0th order phase advanced during a radial period. Since we have already computed it for the 0th order approximation, we denote
\begin{equation}
    \tilde{\varphi} = \int_0^{\chi(t)} f(\chi') d\chi'
\end{equation}
defined for $0<\chi\le2\pi$.

The ratio of $(\dot{\nu}/\nu) dt$ when $\nu$ is a powerlaw in radius is, transformed into terms of $\chi$, equal to $-\alpha e/(2k) \cdot (\sin\chi/(1-e\cos\chi))$. We therefore define, for either cosine or sine,
\begin{equation}
    g_T(\chi) = \int_0^\chi \frac{\sin(\chi')}{1-e\cos(\chi')} T(2\tilde{\varphi}(\chi')) d\chi'.
\end{equation}
These are the integrals that need to be computed numerically for $0<\chi\le2\pi$.

Computing the integral in Equation \ref{eq:1storder} is then a matter of applying the appropriate sin- or cosine- angle sum identity and collecting terms. For $\chi(t_0) < \chi \le 2\pi$,
\begin{equation}
    I_{\sin} = F_{\sin}(\chi) = -\frac{\alpha e}{2k} \big( \sin(2\varphi_\mathrm{IC}) (g_{\cos}(\chi) - g_{\cos}(\chi(t_0))) + \cos(2\varphi_\mathrm{IC}) (g_{\sin}(\chi) - g_{\sin}(\chi(t_0)) \big) \ \mathrm{for}\ \chi(t_0)<\chi\le 2\pi
\end{equation}
Now, in the next radial period, $I_T$ starts at $F(2\pi)$, and the constant terms in $\varphi$ now include a single factor of $\phi_{T_r}$, so we have
\begin{equation}
    I_{\sin} = F_{\sin}(2\pi) - \frac{\alpha e}{2k}\big( \sin(2(\varphi_\mathrm{IC}+\phi_{T_r})) g_{\cos}(\chi-2\pi)  + \cos(2(\varphi_\mathrm{IC}+\phi_{T_r})) g_{\sin}(\chi-2\pi) \big) \ \mathrm{for}\ 2\pi<\chi\le 4\pi
\end{equation}
Each subsequent orbit adds a factor of $\phi_{T_r}$ to the constant-phase terms, as well as a term proportional to each of the $g_T(2\pi)$, so that for $\chi>4\pi$ we have
\begin{eqnarray}
    I_{\sin} = F_{\sin}(2\pi)- \frac{\alpha e}{2k}\bigg( & \sin(2(\varphi_\mathrm{IC}+N_\mathrm{orb}\phi_{T_r})) g_{\cos}(\chi-2\pi N_\mathrm{orb})  + \cos(2(\varphi_\mathrm{IC}+N_\mathrm{orb}\phi_{T_r})) g_{\sin}(\chi-2\pi N_\mathrm{orb}) \\
    &+ g_{\cos}(2\pi) \sum_{k=1}^{N_\mathrm{orb}} \sin(2(\varphi_\mathrm{IC} + k\varphi_{T_r})) + g_{\sin}(2\pi) \sum_{k=1}^{N_\mathrm{orb}} \cos(2(\varphi_\mathrm{IC} + k\varphi_{T_r})) \bigg). \nonumber
\end{eqnarray}
The expression for $I_{\cos}$ has the same factors but rearranged to reflect the cosine, rather than the sine, sum of angles formula. With these integrals in hand, we can then compute $\psi^{(1)}(t) = \varphi(t) + (1/2) I_{\sin}(t)$ and $\mathcal{I}^{(1)} = \mathcal{I}(t_0) \exp( -I_{\cos})$.

It is worth emphasizing that higher-order terms require the numerical computation of (in addition to $g_T$), the 4 integrals of the form
\begin{equation}
    h_{T,U}(\chi) = \int_0^\chi \frac{\sin\chi'}{1-e\cos\chi'} T(2\tilde{\varphi}(\chi')) U(I_{\sin}(\chi')) d\chi',
\end{equation}
where $T$ and $U$ can each be sine or cosine.

\section{Partial derivatives of the time integrand}
\label{sec:taylor}

In section \ref{sec:t} we expand the integrand
\begin{equation}
    f(\chi, e,\nu_k) = (1-e\cos\chi)^{\nu_k} \cos(n\chi)
\end{equation}
as a Taylor series about $e=e_0$, for which we need to know as many derivatives with respect to $e$ as terms we would like to include in the Taylor series. These derivatives are
\begin{equation}
    f_{e^j} = (-1)^j \frac{\Gamma(\nu_k+1)}{\Gamma(\nu_k-j+1)} (1-e_0 \cos\chi)^{\nu_k - j} (\cos\chi)^j \cos(n\chi).
\end{equation}
In addition to these $j$ derivatives with respect to $e$, we would like to be able to compute the mixed partial derivatives $f_{e^j \nu_k^m}$, which denote the $j$th partial derivative of $f$ with respect to $e$ and the $m$th partial derivative with respect to $\nu_k$.

We will write out the first few of these making use of the polygamma functions\footnote{Not to be confused with the other uses of $\psi$ throughout this work: the potential, and the phase of the z-motion in the Volterra Method.} $\psi^{(m)}(z)$ defined as the $m$th logarithmic derivative of the Gamma function $\Gamma(z)$. The first partial derivative with respect to $\nu_k$ of $f_{e^j}$ is
\begin{equation}
\label{eq:Adef}
    f_{e^j\nu_k^1} = f_{e^j}(\chi,e_0,\nu_{k_0}) \left( \psi^{(0)}(\nu_{k_0}+1) - \psi^{(0)}(\nu_{k_0}-j+1) + \ln(1-e_0\cos\chi)\right) \equiv f_{e^j}(\chi,e_0,\nu_{k_0}) A,
\end{equation}
evaluated at $\nu_k=\nu_{k_0}$ and $e=e_0$. We denote the second factor in this equation $A$ and note that its derivatives with respect to $\nu_k$ are
\begin{equation}
\label{eq:Aderiv}
    \frac{\partial^m A}{\partial \nu_k^m} \equiv A^{(m)} = \left( \psi^{(m)}(\nu_k+1) - \psi^{(m)}(\nu_k-j+1) \right) \ \mathrm{for}\ m>0
\end{equation}
The sequence of mixed partial derivatives is then straightforward to compute via application of the product rule and collecting like terms. Each row is always $f_{e^j}$ multiplied by some polynomial function of $A$ and its derivatives $g(A,A^{(1)},A^{(2)},...)$. To go from one line to the next, $g$ is multiplied by $A$ and added to the derivative of $g$ with respect to $\nu_k$ (the clearest example of this is going from the first to the second of the following equations, for which $g(A)=A$).
\begin{eqnarray}
\label{eq:partialsstart}
    f_{e^j\nu_k^1} &=& f_{e^j} A \\
    f_{e^j\nu_k^2} &=& f_{e^j} (A^2 + A^{(1)}) \\
    f_{e^j\nu_k^3} &=& f_{e^j} (A^3 + 3A A^{(1)} + A^{(2)}) \\
    f_{e^j\nu_k^4} &=& f_{e^j} (A^4 + 6A^2 A^{(1)} + 4 A A^{(2)} + 3 (A^{(1)})^2 + A^{(3)} )  \\
    \label{eq:partialsend}
    f_{e^j\nu_k^5} &=& f_{e^j} \left( A^5 + 10A^3 A^{(1)} + 10 A^2 A^{(2)} + 15 A (A^{(1)})^2 + 5 A A^{(3)} +  10 A^{(1)} A^{(2)} + A^{(4)}  \right)
\end{eqnarray}
Keeping in mind that these quantities will be integrated over $\chi$, it is worth emphasizing that $A$, but not $A^{(m)}$ for $m>0$, depends on $\chi$. The integrals that actually need to be computed are therefore
\begin{equation}
\label{eq:integrands}
T_n(\chi, k,e;j,m) = \int_0^\chi (1-e_0 \cos\chi')^{\nu_k - j} (\cos\chi')^j \cos(n\chi') A^m d\chi'
\end{equation}
for values of $m$ and $j$ up to the desired order in the Taylor series. Note that we could define the $T_n$ in terms of $(\ln(1-e\cos\chi))^m$ (the only term in $A$ that depends on $\chi$) rather than $A^m$, but this would introduce an additional sum (see Equation \ref{eq:qn} below), so provided that using $A$ rather than just its logarithmic part does not decrease the accuracy of the integrations, it is better to use $A$. The terms in the Taylor series are said to be $n$th order if $m+j=n$, though it is worth noting that offsets in $e$ will tend to be larger than the offsets in $\nu_k$, so it is likely that fewer orders of $\nu_k$ will be necessary to achieve a given accuracy.

The Taylor series we are evaluating is centered at $e=e_0$ and $\nu_k = \nu_{k_0}$ so that $\delta e = e - e_0$ and $\delta \nu_k = \nu_k - \nu_{k_0}$. At each $\chi$ and $n$, we therefore estimate for arbitrary nearby values of $e$ and $\nu_k$
\begin{equation}
\label{eq:taylor}
    \tau_n(\chi;k,e) = \sum_{j=0}^\infty\sum_{m=0}^\infty \frac{1}{m!j!} (\delta e)^j (\delta \nu_k)^m \int_0^\chi f_{e^j\nu_k^m} (\chi', e_0, \nu_{k_0} ) d\chi',
\end{equation}
so the final task is to write down the integrand in terms of the $T_n$. We can write each of the partial derivatives
\begin{equation}
\label{eq:partialsexp}
    f_{e^j \nu_k^m} =f_{e^j} \sum_{p=0}^m a_p^{(m)} A^p,
\end{equation}
where the $a_p^{(m)}$ can be read off from equations \ref{eq:partialsstart}-\ref{eq:partialsend}. The $m$ superscript emphasizes that the $a_p$ are different for each $m$. For instance,
\begin{eqnarray}
    a_0^{(2)} = A^{(1)}, \ \ \ a_1^{(2)} = 0,\ \ \ a_2^{(2)} = 1 \\
    a_0^{(3)} = A^{(2)}, \ \ \ a_1^{(3)} = 3 A^{(1)},\ \ \ a_2^{(3)} = 0 \ \ \ a_3^{(3)} = 1. \nonumber
\end{eqnarray}

We can now substitute Equation \ref{eq:partialsexp} into Equation \ref{eq:taylor} and use Equation \ref{eq:integrands} to encapsulate the integrals.
\begin{equation}
\label{eq:qn}
    \tau_n(\chi;k,e) = \sum_{j=0}^\infty \frac{(-\delta e)^j}{j!} \frac{\Gamma(\nu_{k_0}+1)}{\Gamma(\nu_{k_0}-j+1)} \sum_{m=0}^\infty \frac{(\delta \nu_k)^m}{m!}  \sum_{p=0}^m a_p^{(m)}  T_n(\chi,k_0,e_0;j,p) 
\end{equation}

To compute the 0th order phase integral for the Volterra method of calculating $z(t)$, we follow exactly the same procedure but compute the Taylor expansion in $\mu_k = \nu_k - \alpha/(2k)$ rather than $\nu_k$. Exactly the same equations as above then apply with $\nu_{k_0} \rightarrow \mu_{k_0}$.

\bibliography{tngAccretion,Epicycles}
\bibliographystyle{aasjournal}

\end{document}